\newcommand{\Swift}{{\it Swift}}
\documentclass{pasj01}
\draft 
\Received{$\langle$reception date$\rangle$}
\Accepted{$\langle$acception date$\rangle$}
\Published{$\langle$publication date$\rangle$}

\begin{document}

\title{ALMA and RATIR observations of GRB131030A}
\author{
Kuiyun Huang\altaffilmark{1},
Yuji Urata\altaffilmark{2}, 
Satoko Takahashi\altaffilmark{3,4},
Myungshin Im\altaffilmark{5},
Po-Chieh Yu\altaffilmark{1},
Changsu Choi\altaffilmark{5},
Nathaniel Butler\altaffilmark{6},
Alan M. Watson\altaffilmark{7}, 
Alexander Kutyrev\altaffilmark{8}, 
William H. Lee\altaffilmark{7}, 
Chris Klein\altaffilmark{9},           
Ori D. Fox\altaffilmark{9,10},                
Owen Littlejohns\altaffilmark{6},       
Nino Cucchiara\altaffilmark{8}, 
Eleonora Troja\altaffilmark{11,8},           
Jes\'{u}s Gonz\'{a}lez\altaffilmark{7},   
Michael G. Richer\altaffilmark{7},     
Carlos Rom\'{a}n-Z\'{u}\~{n}iga\altaffilmark{7},         
Josh Bloom\altaffilmark{9},                            
J.Xavier Prochaska\altaffilmark{12},                    
Neil Gehrels\altaffilmark{8},                    
Harvey Moseley\altaffilmark{8},                  
Leonid Georgiev\altaffilmark{7},       
Jos\'{e} A. de Diego\altaffilmark{7},      
Enrico Ramirez‐Ruiz\altaffilmark{12}                      
}

%
\altaffiltext{1}{Department of Mathematics and Science, National Taiwan Normal University, Lin-kou District, New Taipei City 24449, Taiwan} 
\altaffiltext{2}{Institute of Astronomy, National Central University, Chung-Li 32054, Taiwan}
\altaffiltext{3}{Joint ALMA Observatory, Alonso de Cordova 3108, Vitacura, Santiago, Chile}
\altaffiltext{4}{National Astronomical Observatory of Japan, 2-21-1 Osawa, Mitaka, Tokyo 181-8588, Japan}
\altaffiltext{5}{Center for the Exploration of the Origin of the Universe, Department of Physics \& Astronomy, FPRD, Seoul National University, Shillim-dong, San 56-1, Kwanak-gu, Seoul, Korea}
\altaffiltext{6}{School of Earth \& Space Exploration, Arizona State University, Tempe, AZ 8528}
\altaffiltext{7}{Instituto de Astronom\'{i}a, Universidad Nacional Aut\'{o}noma de M\'{e}xico, Apartado Postal 70-264, 04510 M\'{e}xico, D. F., M\'{e}xico}
\altaffiltext{8}{NASA, Goddard Space Flight Center, Greenbelt, MD 20771}
\altaffiltext{9}{Astronomy Department, University of California, Berkeley, CA 94720-7450}
\altaffiltext{10}{Space Telescope Science Institute, 3700 San Martin Drive, Baltimore,
  MD 21218}
\altaffiltext{11}{Department of Astronomy, University of Maryland, College Park, MD 20742}
\altaffiltext{12}{Department of Astronomy and Astrophysics, UCO/Lick Observatory, University of California, 1156 High Street, Santa Cruz, CA 95064}

\email{urata@astro.ncu.edu.tw}

\KeyWords{gamma-ray burst: individual (GRB131030A) --- Submillimeter: galaxies --- X-rays: bursts}

\maketitle

\begin{abstract}

  We report on the first open-use based Atacama Large Millimeter/submm
  Array (ALMA) 345-GHz observation for the late afterglow phase of
  GRB131030A. The ALMA observation constrained a deep limit at 17.1 d
  for the afterglow and host galaxy. We also identified a faint
  submillimeter source (ALMAJ2300-0522) near the GRB131030A position.
  The deep limit at 345 GHz and multifrequency observations obtained
  using {\it Swift} and RATIR yielded forward shock modeling with a
  two-dimensional relativistic hydrodynamic jet simulation and
  described X-ray excess in the afterglow. The excess was inconsistent
  with the synchrotron self-inverse Compton radiation from the forward
  shock. The host galaxy of GRB131030A and optical counterpart of
  ALMAJ2300-0522 were also identified in the SUBARU image.  Based on
  the deep ALMA limit for the host galaxy, the 3-$\sigma$ upper limits
  of IR luminosity and the star formation rate (SFR) is estimated as
  $L_{IR}<1.11\times10^{11} L\solar$ and SFR$<18.7$
  ($M\solar$~yr$^{-1}$), respectively. Although the separation angle
  from the burst location ($3\farcs5$) was rather large,
  ALMAJ2300-0522 may be one component of the GRB131030A host galaxy,
  according to previous host galaxy cases.

\end{abstract}

\section{Introduction}

Submillimeter (submm) and millimeter (mm) follow-up observations have
played an essential role in identifying gamma-ray burst (GRB)
afterglow and host galaxies in, for example delineating the energy
scale, geometry, radiation physics, and environments of long GRBs
(e.g. \cite{frail02, sheth03, urata14}). However, submm/mm follow-up
observations have lagged behind X-ray, optical and cm radio
observations (summaries of afterglow observations are available in
\cite{prealma,submmgrb2}) because of the limited sensitivity of
previous submm/mm facilities coupled with the higher redshift of
\textit{Swift} GRBs.

The Atacama Large Millimeter/submm Array (ALMA) was first used in the
early science phase for GRB host galaxies, and its observations have
provided exceptional results \citep{wang12,hatsukade,berger14}.
However, observations for afterglow phase are still limited
because of the observation guidelines of ALMA (e.g. a 3-week reaction
time since ToO triggering) for the early science phase (e.g. Cycle 1).
Here, we report the first open-use based ALMA observation of the late
afterglow phase of GRB131030A.

GRB 131030A was detected using the \textit{Swift} \citep{swiftsat}
Burst Alert Telescope (BAT) at 20:56:18 UT on 2013 October 30
\citep{gcn15402}. The duration, $T_{90}$ in the 15-350 keV band was
$41.1\pm4.0$ s \citep{gcn15456}. The afterglow in X-ray and optical
bands was also identified using the \textit{Swift} X-ray Telescope
(XRT) and Ultraviolet/Optical Telescope (UVOT). On the basis of the
UVOT observation, the burst position was determined at $23^{\rm
  h}00^{\rm m}16^{\rm s}.14$, $-05^{\circ}22'05'.2$ with a 90\%
confidence error radius of $0\farcs5$ \citep{gcn15414}. The redshift
of GRB 131030A was measured at $z=1.293$ on the basis of the optical
spectroscopic observation by using the Nordic Optical Telescope
\citep{gcn15407}.
The polarized early (655s to 2 hrs) optical lightcurve of the afterglow
was also observed using the RoboPol instrument \citep{pol}.
The Konus-Wind observation also revealed a prompt emission and
characterized the spectrum properties in the 20 keV $-$ 15 MeV
range. The time-averaged spectrum was fitted using the Band function
with the spectrum peak energy $E^{obs}_{peak}$ of 177$\pm$10 keV. The
isotropic energy $E_{iso}$ was also estimated as
$(3.0\pm0.2)\times10^{53}$ erg, assuming cosmological parameters of
$H_{0}=$ 70 ${\rm km}$ ${\rm s^{-1}}$ ${\rm Mpc^{-1}}$, $\Omega_m=0.27$, and
$\Omega_{\Lambda}=0.73$ \citep{gcn15413}.

\section{Observations and Results}

\subsection{ALMA Submillimeter Follow-up Observation}

\begin{figure}
 \begin{center}
  \includegraphics[width=17cm]{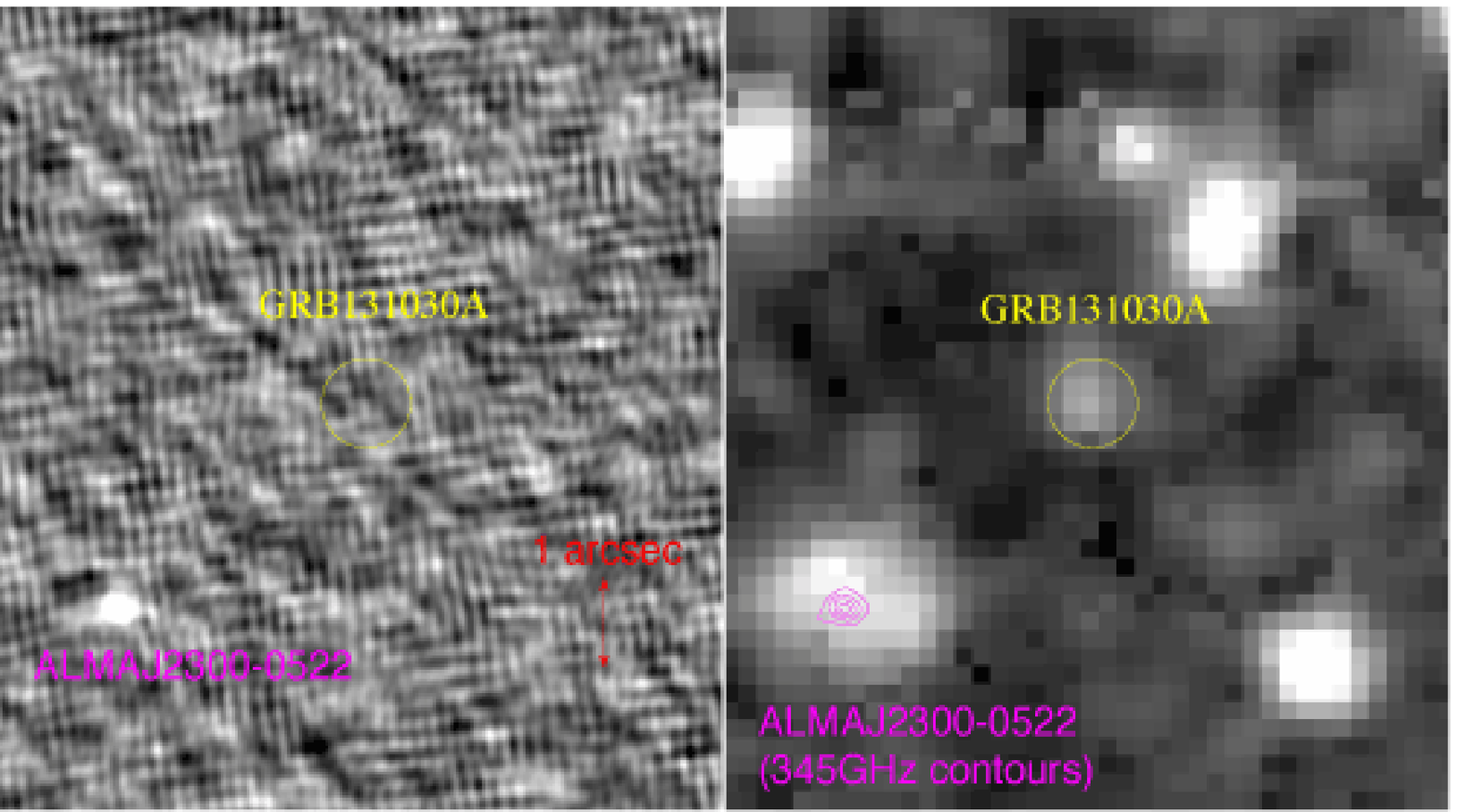}
 \end{center}
\caption{Images of the GRB 131030A field in the 345-GHz band that were obtained using ALMA
  (left) and in the optical $Rc$-band that were obtained using SUBARU (right). The host
  galaxy of GRB 131030A was detected in the $Rc$-band image. The submm
  source is located $3\farcs5$ away from the GRB position with an optical
  counterpart. The magenta lines in the right panel display the contour (start from 2.5 $\sigma$ with 2.5 $\sigma$ step) of the ALMA image.\label{host}}
\end{figure}

\begin{figure}
 \begin{center}
  \includegraphics[width=16cm]{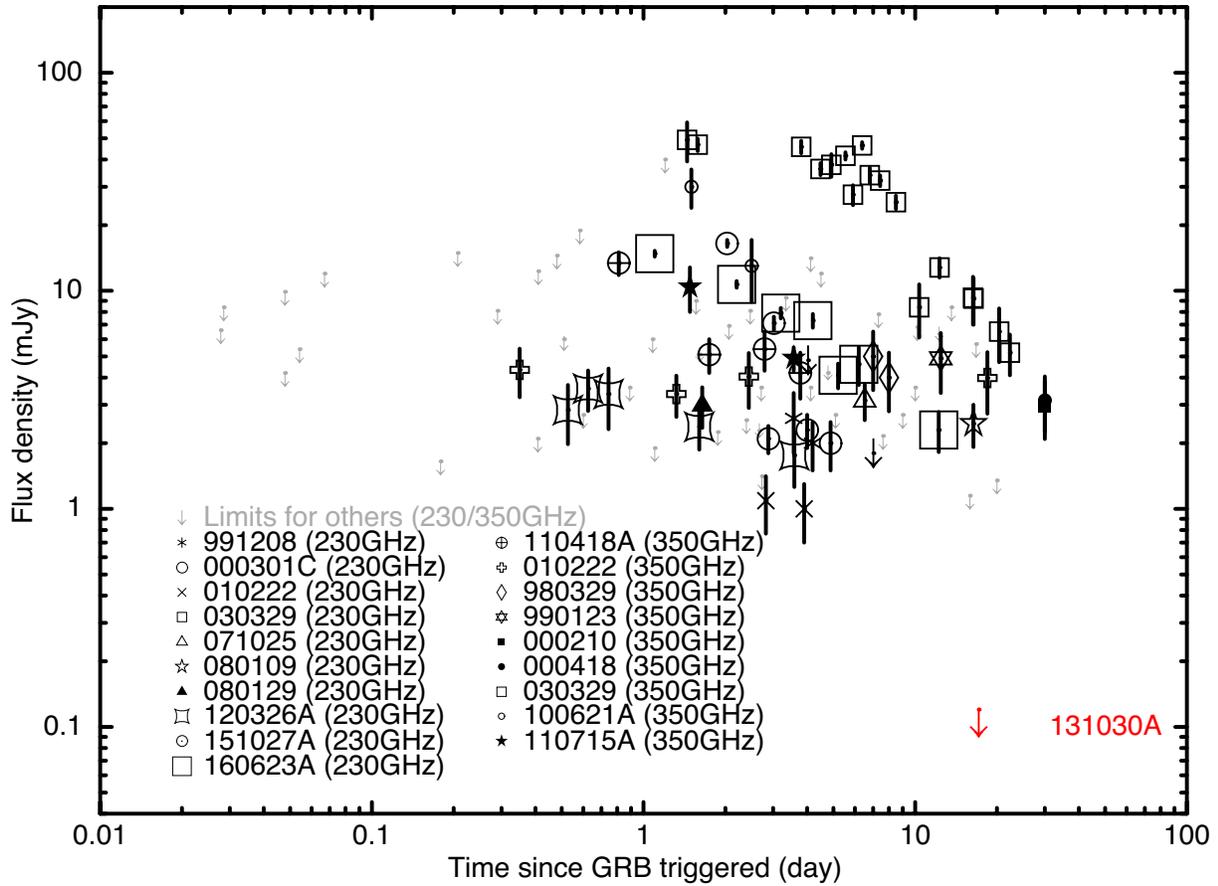}
 \end{center}
 \caption{Light curve summary of afterglow observations in submm bands (230 and 345 GHz). The ALMA observation for GRB131030A is significantly deep among other observations. \label{submmaf}}
\end{figure}

We used ALMA to observe the afterglow at 345-GHz (default
  continuum setup) with the C32-5 configuration under the Cycle
  1 open-use mode. Although the observation policy of the ALMA Cycle
1 includes a 3-week reaction time restriction for its execution
through ToO triggering, the 345-GHz observation was initiated at 23:11
UT on 2013 November 16 (17.12 d after the GRB).
Thus, the observation was executed several days earlier than expected.
Three calibrators (J2232+117, J2148+0657, and J2301$-$0158) for flux,
bandpass, and phase calibrations, respectively, were also observed.
The on-source time and total observing time were 53 min and 90 min, respectively.
The raw data were calibrated using the Common Astronomy Software
Applications 4.1 (CASA, \cite{casa}) with the standard procedure, and
final CLEANed images were made using the ``clean'' task with a robust
briggs weighting (robust parameter of 0.5). The resulting
synthesized beam sizes were $0\farcs252\times0\farcs207$ with a
position angle of 68.2 deg.
No source was observed from the location of the optical afterglow to
the 3$-\sigma$ limit of 0.12 mJy (Figure \ref{host} left). As shown in
Figure \ref{submmaf}, this limit is significantly deep in the submm
bands among other afterglow observations. Although ALMA observed the
bright submm afterglow associated with GRB110715A ($\sim5$ mJy at 3.6
d) during the commissioning phase \citep{prealma}, the quality of
observation was comparable with those of other smaller submm
instruments. Hence, this demonstration of the high sensitivity
follow-up would be a benchmark with further submm observations.
As shown in Figure \ref{host}
(left), a source (ALMAJ2300-0522) with $0.716\pm0.045$ mJy was
identified in the ALMA image at $23^{\rm h}00^{\rm m}16^{\rm s}.326$,
$-05^{\circ}22'07'.50$ (approximately $3\farcs5$ from the GRB position).
Although several astrometry-related bugs (e.g. phase calibrator
coordinate inconsistency) were reported from the ALMA ARC after data
delivery, we could not reasonably explain the significant offsets of
approximately $3\farcs5$ (more than 10-fold the beam size) from
the GRB position. In addition, the source also had an optical
counterpart in the SUBARU \textit{Rc}-band image (Figure \ref{host} right,
$\S$2.5).

\subsection{Optical and Near Infrared Afterglow Follow-ups}

We used the Reionization and Transients InfraRed camera (RATIR) to monitor the afterglow in \textit{r-}, \textit{i-},
\textit{Z-}, \textit{Y-}, \textit{J-} and \textit{H-}bands from
1.763$\times10^{4}$ s to 7.220$\times10^{5}$ s after the burst.
RATIR is a six band simultaneous optical and NIR imager mounted on the
autonomous 1.5~m Harold L. Johnson Telescope at the Observatorio
Astron\'{o}mico Nacional on Sierra San Pedro M\'{a}rtir in Baja
California, Mexico
\citep{2012SPIE.8446E..10B,2012SPIE.8444E..5LW,2012SPIE.8453E..2SK,2012SPIE.8453E..1OF}.
%
%
The images were reduced in near real-time using an automatic
pipeline. Bias subtraction and twilight flat division were performed
using algorithms written in {\sc python}, image alignment was conducted
by astrometry.net \citep{2010AJ....139.1782L} and image co-addition was
achieved using {\sc swarp} \citep{2010ascl.soft10068B}.\par

We performed photometry for individual science frames and mosaic using
{\sc sextractor} \citep{1996A&AS..117..393B} with apertures ranging
from 2 to 30 pixels ($0\farcs64-9\farcs6$ in optical,
$0\farcs6-9\farcs0$ in NIR). Based on the weighted average of the flux
in these apertures for all stars in a field, an annular
point-spread-function (PFS) was constructed.  We then optimized point
source photometry by fitting this PSF to the annular flux values of
each source.  The photometric calibration in \textit{r-}, \textit{i-},
\textit{Z-}, \textit{J} and \textit{H} bands were made comparing with
the Sloan Digital Sky Survey Data Release 9 (SDSS) and the Two Micron
All Sky Survey.  The RATIR and SDSS \textit{r-}, \textit{i-} and
\textit{Z-}bands agree to within $\lesssim$3 per cent (Butler et
al. 2016 in prep.).  For the \textit{Y-}band calibration, we used an
empirical relation in terms of \textit{J} and \textit{H} magnitudes
derived from the United Kingdom Infrared Telescope (UKIRT) Wide Field
Camera observations \citep{2009MNRAS.394..675H,2007A&A...467..777C}.

We also conducted the \textit{B-} and \textit{R-}band follow-up observations by
using the robotic 1-m telescope at the Mt. Lemmon observatory, which
is operated by the Korea Astronomy Space Science Institute
\citep{lee,han}. The observations were initiated at 01:43 UT on 2013 October 31 (1.720$\times10^{4}$ s after the burst), and three epochs for monitoring were
conducted during the same run.
A standard routine including bias subtraction and flat-fielding
corrections was employed to process the data using the {\sc IRAF}
package. The {\sc DAOPHOT} package was used to perform aperture photometry
of the GRB images. For the photometric calibration of the afterglow,
several stars from the NOMAD catalog \footnotemark were chosen.
To remove the effects of the Galactic interstellar extinction, we used
the reddening map by \citet{schlafly}.
\footnotetext{http://www.usno.navy.mil/USNO/astrometry/optical-IR-prod/nomad}

\subsection{Afterglow Lightcurve}

The multifrequency lightcurve is shown in Figure \ref{lc} (left). 
%
%
The X-ray data were obtained from the online data repository prepared
by \citet{butler07}.
The temporal evolution in the \textit{r-} and \textit{i-}bands indicates an
achromatic temporal break at approximately $\sim2\times10^{5}$ s.
In contrast, X-ray light curve later than $\sim4\times10^{3}$ sec shows
the simple evolution.
To describe the light curves, we employed a single power-law function by
dividing light curve data into earlier ($t<\sim2\times10^{5}$) and
later ($t>\sim2\times10^{5}$) phases.
We successfully fitted the single power-law function to the earlier
light curves.  The \textit{r-}, \textit{i-}, and X-ray band light
curves in the later phase were also described using the single power
law function. The X-ray light curve later than $\sim4\times10^{3}$ sec
was also described with the single power-law function with the decay
index of $\alpha_{X}=-1.25\pm0.02$. Here, we use a notation $F_{\nu}\propto t^{\alpha}\nu^{\beta}$ with sub-indices of observed band. The fitting results are summarized
in Table \ref{tbl-1}.
%
We also successfully fitted the broken power-law model to the \textit{r-} and \textit{i-}bands light curves.
For the \textit{r-}band, we obtained
$\alpha_{r1}=-0.86\pm0.04$,
$\alpha_{r2}=-2.06\pm0.16$, and 
$t_{br}=(2.51\pm0.48)\times10^{5}$ s;
for the \textit{i-}band, we obtained $\alpha_{i1}=-0.82\pm0.04$,
$\alpha_{i2}=-2.04\pm0.17$, and $t_{bi}=(2.51\pm0.48)\times10^{5}$ s.
Here, sub-indices of 1 and 2, and $t_{b}$ indicate before and after
the temporal break, and the break time, respectively. The decay index
of X-ray afterglow before the jet break is also estimated as
$\alpha_{X1}=-1.31\pm0.04$.

\subsection{Afterglow Spectrum}

\begin{figure}
 \begin{center}
  \includegraphics[width=8cm]{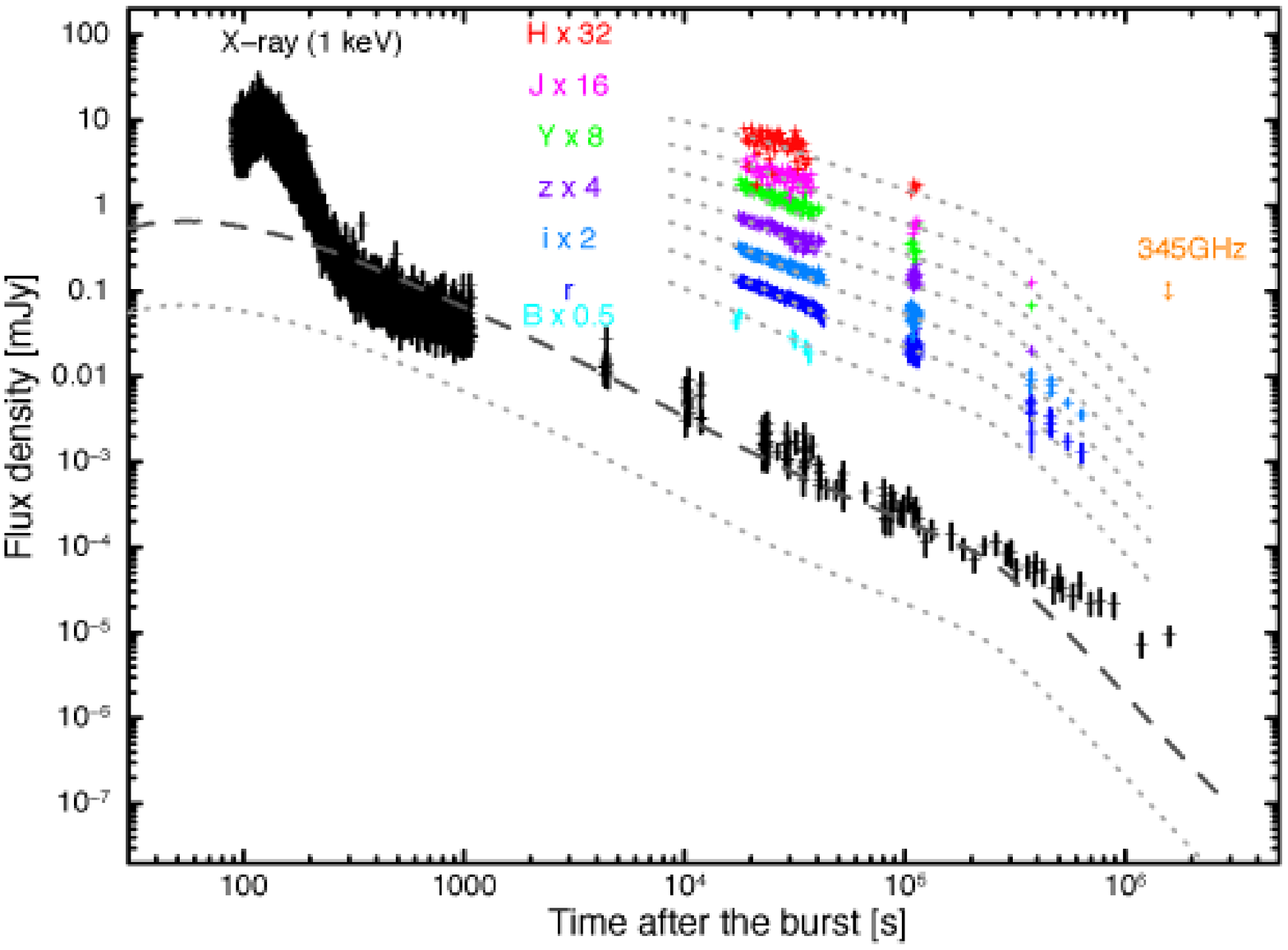}
  \includegraphics[width=8cm]{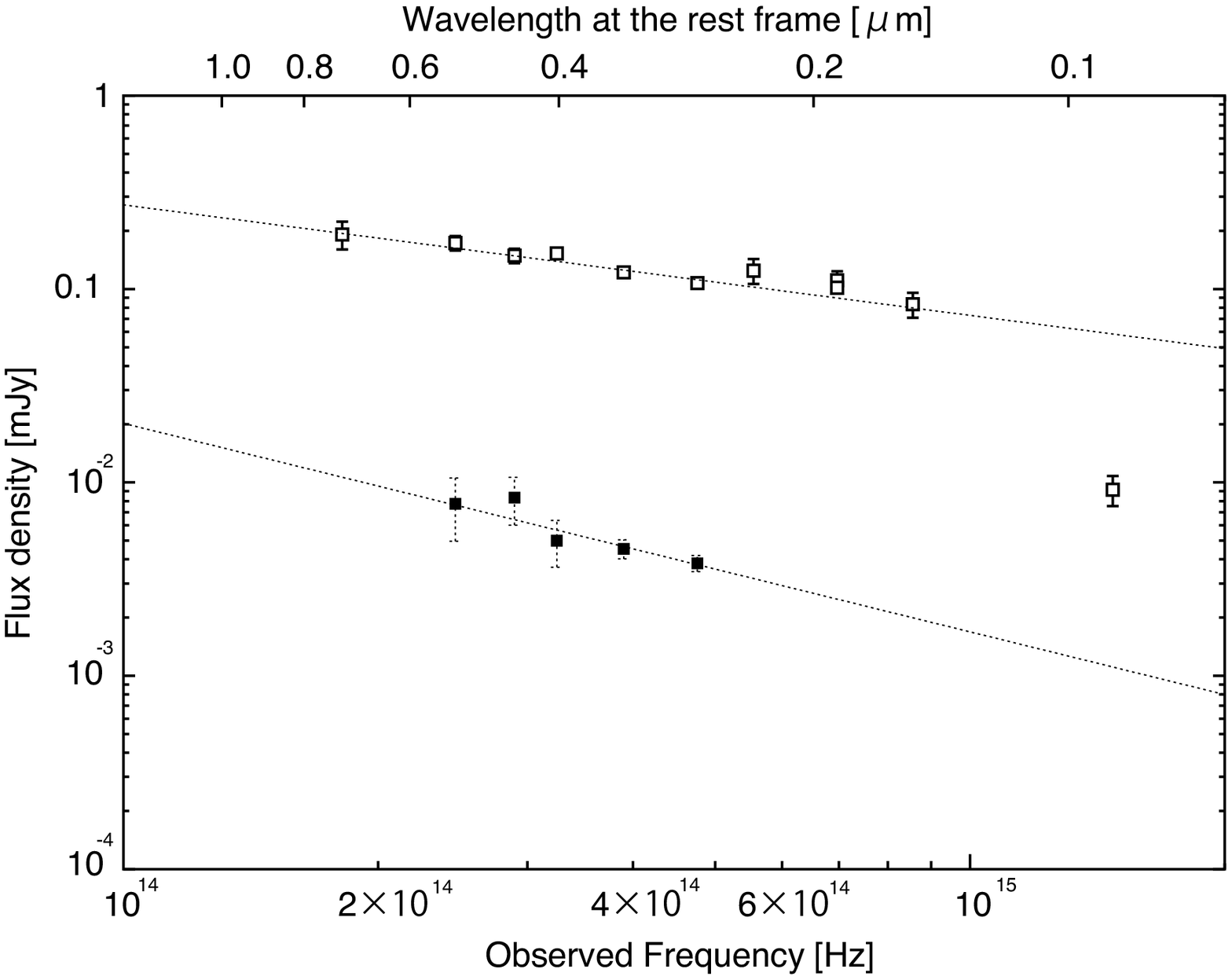}
 \end{center}
\caption{Left panel shows X-ray, optical and submillimeter light curves of the GRB 131030A afterglow. The light-gray dotted lines indicate the optimal modeling functions obtained through the numerical simulation by the boxfit code. The dark-grey dashed line indicates the model function including the SSC component. The right panel shows the SED of the afterglow at $2.32\times10^{4}$ s (0.268 d; open squares) and 3.76$\times10^{5}$ s (4.34 d; filled squares).\label{lc}}
\end{figure}

We generated spectral flux distributions at $2.32\times10^{4}$ s
(0.268 d; first epoch) and 3.76$\times10^{5}$ s (4.34 d; second
epoch). For the first epoch, we also added \textit{UVW2-},
\textit{U-}, \textit{B-}, and \textit{V-}band data that were obtained
using \textit{Swift}/UVOT at approximately 2.3$\times10^{4}$ s after
the burst, which coincided with the RATIR observations.
We used the standard procedure for the UVOT analysis \footnotemark.
As shown in Figure \ref{lc} (right), the sharp spectral drop at approximately
1$\times10^{15}$ Hz that was caused by the Ly$\alpha$ absorption at
$z=1.293$ was identified.  Thus, we fitted the first epoch SED with a
power-law function by excluding the \textit{UVW2} data and obtained
$\beta=-0.57\pm0.09$ ($\chi^{2}/\nu=1.9$ with $\nu=8$) for the first
epoch and $\beta=-1.08\pm0.23$ ($\chi^{2}/\nu=0.34$ with $\nu=3$) for
the second epoch.  As shown in Figure \ref{lc} (right), the
optical SED, including the UV data points obtained using UVOT, were
effectively described by the single power-law function without
considering the host galaxy extinction.
\footnotetext{http://www.swift.ac.uk/analysis/uvot/index.php}

\subsection{SUBARU Archive Image and Host Galaxy}

We reduced the SUBARU Suprime-Cam images obtained from the SMOKA
archive system \citep{smoka}. The GRB131030A field was observed using
the $Rc$-bands during the previous follow-up observations of XRF040916
on September 20, 2004. Owing to the wide field of view of the
Suprime-Cam \citep{sc}, the location of GRB 131030A had an effective
exposure time of 2400 s.  The basic reduction of the Suprime-Cam data
was performed using the SDFRED \citep{sdfred}.  The final coadded
image was astrometrically aligned relative to the SDSS catalog,
resulting in an rms scatter of $0\farcs08$.  Figure \ref{host} (right)
provides the $Rc$-band images for the GRB131030A field, and an
extended source was identified inside the position error region
through \Swift/UVOT. By using IRAF centroid, the location of the
source was determined at $23^{\rm h}00^{\rm m}16^{\rm s}.146$,
$-05^{\circ}22'05'.18$, which is $0\farcs07$ away from the afterglow
location. The projected offset of 0.6 kpc is smaller than values for
long GRBs (e.g. \cite{bloom}).  Thus, we concluded that the source was
the host galaxy of GRB131030A.  The brightness with a $1\farcs0$
radius was $Rc=26.23\pm0.11$.  The optical counterpart of
ALMAJ2300-0522 was also measured as $Rc=24.48\pm0.03$ with a
$1\farcs6$ radius.

\section{Discussions}
\subsection{Forward Shock Radiation and X-Ray Excess}

The multicolor monitoring observations performed using RATIR provided
an effective dataset for describing the forward shock synchrotron
radiation.
The spectral index change between first and second epochs could be
due to the cooling frequency passage in the optical band.
The absence of the sharp cooling break in the optical light curves
could be explained by the smooth transition of spectral regimes
(e.g. \cite{uhm14}).
The closure relation (e.g. \cite{sari}) also indicated that
the optical afterglow at $2.32\times10^{4}$ s and 1.56$\times10^{5}$ s
was consistent with $\nu_{m}<\nu_{opt}<\nu_{c}$ under the interstellar
medium (ISM) with a slow cooling condition
($\alpha=3/2\beta=-0.86\pm0.14$) and jet phase with $nu_{opt}>\nu_{c}$
($\alpha=2\beta=-2.2\pm0.5$), respectively.
Although the optical afterglow showed the jet
break, the X-ray afterglow revealed no temporal break at the jet
break time. The $\alpha_{o}-\alpha_{x}$ relation of $0.44\pm0.06$ for
before the jet break was also the explicit outlier
(e.g. \cite{normal}). Thus, these results indicate that the X-ray
afterglow could have an origin that was distinct from that of the
optical afterglow (e.g. \cite{huang07,troja07}).

\begin{figure}
 \begin{center}
  \includegraphics[width=16cm]{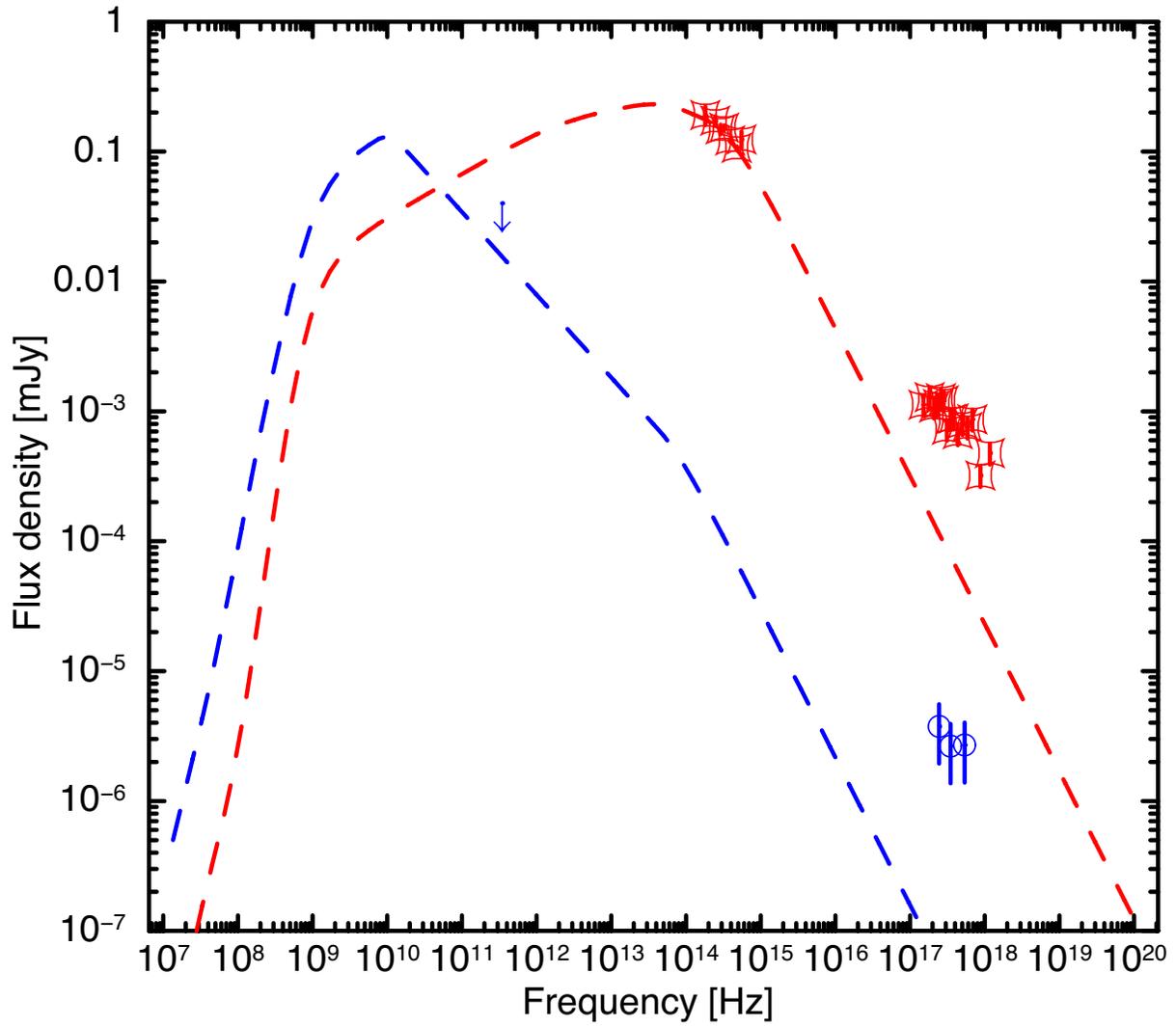}
 \end{center}
\caption{Spectrum energy distribution at 0.2662 d (red) and 17.125 d
  (blue) after the burst. The dashed lines represent the forward shock
  synchrotron model spectrum that was calculated using the boxfit code. The ALMA
  1-$\sigma$ upper limit at 17.125 d is indicated by the blue
  arrow. \label{bsed}}
\end{figure}

To describe the X-ray afterglow behavior, we performed forward shock
synchrotron radiation modeling with the optical light curves and ALMA
upper limit by using the boxfit code \citep{boxfit}. Because the code
involved a two-dimensional relativistic hydrodynamical jet simulation
with a homogeneous circumburst medium, we hereafter, consider only the
ISM condition.  We determined the optimal modeling parameters with the
on-axis case (observing angle, $\theta_{obs}=0$) as
$\theta_{jet}=8^{\circ}.8$, $E=1.05\times10^{52}$ erg,
$n=2.54\times10^{-1}$ cm$^{\rm -3}$, $p$=2.28,
$\epsilon_{B}=4.36\times 10^{-2}$, and
$\epsilon_{e}=2.69\times10^{-1}$.  The values were updated from those
in \citet{xrf} by using data from six OIR bands and the ALMA
1-$\sigma$ limit. Figure \ref{lc} shows the most effective model
functions for multicolor light curves. The broadband SED modeling
results are also provided in Figure \ref{bsed}. These results clearly
show the excess in the X-ray band.

\begin{figure}
 \begin{center}
  \includegraphics[width=16cm]{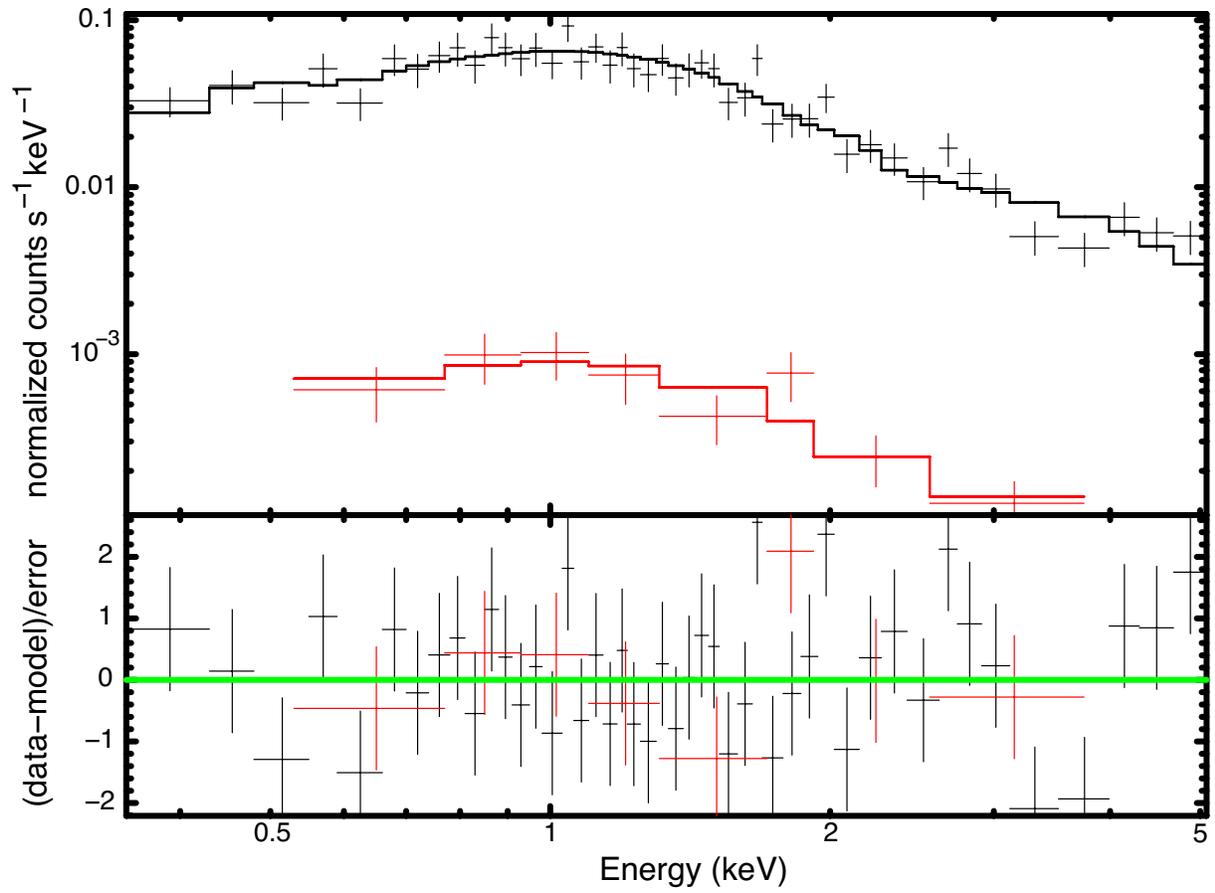}
 \end{center}
 \caption{X-ray spectrum in the time range of $1\times10^{4}-1\times10^{5}$ s (black cross) and $>5\times10^{5} s$ (red cross). The bottom panel shows residual for fitting with power law.\label{xrayspec}}
\end{figure}

On the basis of the analytic solution described by \citet{fan08}, we
calculated the expected flux density of synchrotron
self-inverse-Compton radiation (SSC) by using forward shock modeling.
The reasonable flux scale and differences (approximately 10-fold
times) between synchrotron and SSC for the earlier afterglow were
estimated using the model. However, the SSC model could not explain
the different temporal evolution in the late phase.  In addition, the
expected $\nu^{SSC}_{m}$ pass in the X-ray band ($\sim 4\times10^{5}$
sec) with the analytic solution was not identified in the observed
X-ray spectrum (Figure \ref{xrayspec}).  Hence, SSC is
unlikely to explain the observational properties.
One of the alternative models for describing the distinct temporal
evolution between optical and X-ray is the late prompt emission
\citep{lateprompt}.
A spinning-down magnetar as the central engine is one of the
explanations for the X-ray excess \citep{zhang01, troja07}.


%

\subsection{Host Galaxy and ALMAJ2300-0522}

The ALMA observation also provides a unique upper limit of 345-GHz for
estimating the rest-frame infrared luminosity and star formation rate
(SFR) of the host galaxy.
%
Following \citet{wang12}, we applied a redshifting to a
template infrared SED, which we obtained from the library provided by
\citet{ce01}.
The SED library is
luminosity-dependent ($2\times10^8$ to $4\times10^{13} L\solar$) based
on a locally calibrated luminosity-dust temperature relation and does
not enable scaling the SEDs. We constrained the 3-$\sigma$ upper
limits of IR luminosity as $L_{IR} <1.11\times10^{11} L\solar$.  Using
the SFR conversion of star-forming galaxies, SFR($M\solar$
yr$^{-1})=1.7\times10^{-10} L_{IR}/L\solar$ \citep{kennicutt98} used in
\citet{wang12}, we obtained the 3-$\sigma$ upper limit of SFR as $<18.7$
($M\solar$~yr$^{-1}$).
The upper limit of SFR was consistent with those of the TOUGH samples
that constrained the average SFR as lower than 15 $M\solar$~yr$^{-1}$
\citep{hjorth12,michalowski12}.

\begin{figure}
 \begin{center}
  \includegraphics[width=16cm]{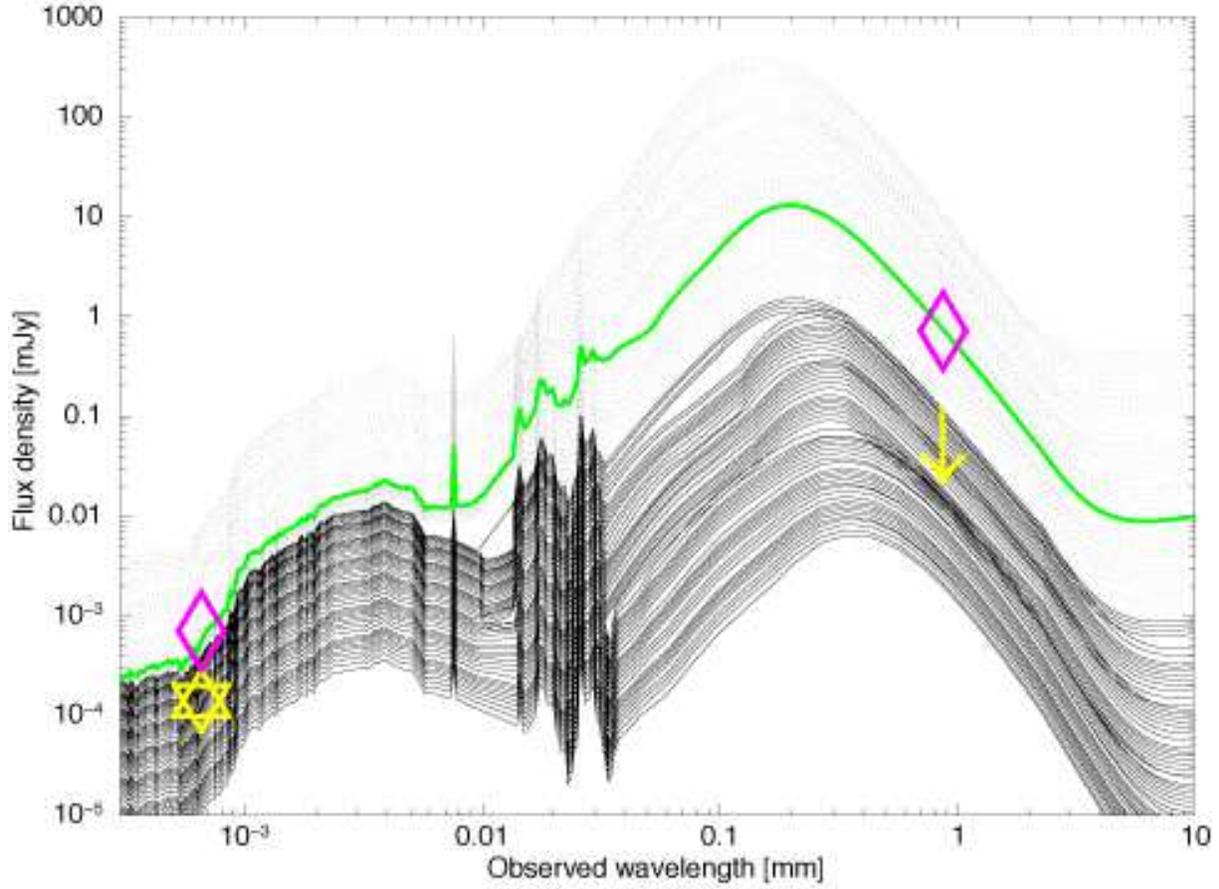}
 \end{center}
\caption{SED of the GRB host galaxy and ALMAJ2300-0522.  ALMA and
  SUBARU observations are indicated with yellow marks for the GRB host
  galaxy and purple diamonds for ALMAJ2300-0522.  The ALMA observation
  constrains the infrared SEDs of the GRB host galaxy with the SED
  templates. The black thick curves represent the templates that
  satisfy the ALMA 3-$\sigma$ upper limit for the GRB host galaxy. The
  green solid line indicates the optimal template for ALMAJ2300-0522,
  on the basis of assuming the same redshift as that of
  GRB131030A.\label{hostsed}}
\end{figure}

ALMAJ2300-0522 is $3\farcs5$  (30.2 kpc) from the GRB
location.
Some of the GRB host galaxies
show companions and exceptionally extended emission.
HST images resolved companions for GRB021004 at $0\farcs28$ (2.4 kpc)
from the host location \citep{021004} , for GRB050820A at $1\farcs3$ (10.7 kpc) \citep{050820a} and
$0\farcs4$ (3.3 kpc), and for GRB080605 at $1\farcs0$ (8.7 kpc) \citep{080605},
respectively.
Most notably, the Ly$\alpha$ imaging for the GRB000926 host galaxy
revealed extended emission of approximately 33.7 kpc at the long axis
\citep{fynbo02}. Thus, ALMAJ2300-0522 is potentially a component of
the GRB131030A host galaxy. By assuming the same redshift with
GRB131030A, we also obtained $L_{IR}=8.65\times10^{11} L\solar$ and
the corresponding SFR of 147 $M\solar$~yr$^{-1}$ by using the same
template redshifting method (green solid line in Figure
\ref{hostsed}).  The obtained values were comparable or lower than
those of observed ultra luminous infrared galaxy host cases
(e.g. \cite{berger01,frail02}).

Another noteworthy feature is that ALMAJ2300-0522 is categorized as a
faint ($<$1 mJy) submm galaxy (SMG). Faint SMGs tend not to have
optical counterparts, in contrast to ALMAJ2300-0522, even when
ultra-deep optical survey data are used
\citep{chen14,ono14,fujimoto16}.
%
%
%
Hence, future studies that use redshift estimation can clarify the
SMG-GRB association.

\section{Conclusion and Summary}
We conducted the first open-use based ALMA observation for the
afterglow of GRB131030A.
The high sensitivity of the ALMA observation would make a significant
improvement in further afterglow follow-ups such as coordinated submm
observations for radio polarimetry and longterm multi-frequency
monitoring.
%
With the comprehensive optical and near infrared afterglow
observations performed through RATIR, we described the forward shock
synchrotron radiation and X-ray excess in the afterglow. The excess is
inconsistent with the SSC model and requires another component such as
late prompt emission.
Our ALMA observation also constrains the 3-$\sigma$ limit of infrared
luminosity and SFR of the host galaxy, which is consistent with other
nearby $z<\sim1$ samples. The submm source, ALMAJ2300-0522 is located
$3\farcs5$ from the GRB location. Although the separation is rather large,
the source may be one component of the host galaxy, according to
previous host galaxy cases (e.g. GRB000926). To further analyze the
association between ALMAJ2300-0522 and GRB131030A, secure redshift
estimation is required.

\begin{ack}
This paper makes use of the following ALMA data:
ADS/JAO.ALMA\#2012.1.00875.T. ALMA is a partnership of ESO
(representing its member states), NSF (USA) and NINS (Japan), together
with NRC (Canada), NSC and ASIAA (Taiwan), and KASI (Republic of
Korea), in cooperation with the Republic of Chile. The Joint ALMA
Observatory is operated by ESO, AUI/NRAO and NAOJ.
We also thank EA-ARC, especially Shigehisa Takakuwa and Trejo-Cruz
Alfonso for various supports on ALMA observation and reduction.
This work is partly supported by the Ministry of Science and
Technology of Taiwan grants MOST 104-2112-M-008-011 and 105-2112-M-008-013-MY3 (YU).
We thank the RATIR project team and the staff of the Observatorio
Astron\'{o}mico Nacional on Sierra San Pedro M\'{a}rtir. RATIR is a
collaboration between the University of California, the Universidad
Nacional Auton\'{o}ma de M\'{e}xico, NASA Goddard Space Flight Center,
and Arizona State University, benefiting from the loan of an H2RG
detector and hardware and software support from Teledyne Scientific
and Imaging. RATIR, the automation of the Harold L. Johnson Telescope
of the Observatorio Astron\'{o}mico Nacional on Sierra San Pedro
M\'{a}rtir, and the operation of both are funded through NASA grants
NNX09AH71G, NNX09AT02G, NNX10AI27G, and NNX12AE66G, CONACyT grants
INFR-2009-01-122785 and CB-2008-101958 , UNAM PAPIIT grant IG100414
and CONACyT LN260369, and UC MEXUS-CONACyT grant CN 09-283.
MI and CC acknowledge the support from the National Research
Foundation of Korea grant, No. 2008-0060544, funded by the Korea
government (MSIP).
This work made use of data supplied by the UK Swift Science Data
Centre at the University of Leicester.

\end{ack}

\clearpage

\begin{table}
\begin{center}
\caption{Summary of lightcurve fitting.\label{tbl-1}}
\begin{tabular}{ccccc}
\hline
\hline
\multicolumn{5}{c}{Single PL}\\
\cline{2-5}
& \multicolumn{2}{c}{Earlier phase} & \multicolumn{2}{|c}{Later phase}\\
& \multicolumn{2}{c}{($1\times10^{4}<t<2\times10^{5}$ s)} & \multicolumn{2}{|c}{($t>\sim2\times10^{5}$ s)}\\
Filter & $\alpha$ & $\chi^{2}/\nu(\nu)$ & \multicolumn{1}{|c}{$\alpha$} & $\chi^{2}/\nu(\nu)$ \\
\textit{B} & $-1.17\pm0.15$ & 2.8  (4)   & \multicolumn{1}{|c}{...}& ... \\
\textit{r} & $-1.00\pm0.01$ & 1.58 (307) & \multicolumn{1}{|c}{$-2.08\pm0.33$} & 1.04 (11) \\
\textit{i} & $-0.99\pm0.01$ & 1.37 (300) & \multicolumn{1}{|c}{$-2.07\pm0.36$} & 0.83 (6) \\
\textit{z} & $-0.97\pm0.02$ & 1.87 (130) & \multicolumn{1}{|c}{...}& ... \\
\textit{Y} & $-0.95\pm0.02$ & 1.25 (106) & \multicolumn{1}{|c}{...}& ... \\
\textit{J} & $-0.96\pm0.04$ & 1.79 (92)  & \multicolumn{1}{|c}{...}& ... \\
\textit{H} & $-0.94\pm0.07$ & 2.02 (79)  & \multicolumn{1}{|c}{...}& ... \\
X-ray      & $-1.31\pm0.04$ & 0.97 (94)  & \multicolumn{1}{|c}{$-1.29\pm0.10$} & 0.78 (18)\\
\hline
\hline
& \multicolumn{2}{c}{($t>4\times10^{3}$ s)} & \multicolumn{2}{|c}{}\\
X-ray  &  $-1.25\pm0.02$ & 0.91 (125) &  \multicolumn{1}{|c}{...}  & ...\\
\hline
\hline
\multicolumn{5}{c}{Broken PL}\\
\cline{2-5}
Filter & $\alpha_{1}$ & $\alpha_{2}$ & $t_{b} (s)$ & $\chi^{2}/\nu$ \\
\hline
\textit{r} & $-0.86\pm0.04$ & $-2.06\pm0.16$ & $(2.51\pm0.48)\times10^{5}$ & 1.41 (316)\\
\textit{i} & $-0.82\pm0.04$ & $-2.04\pm0.17$ & $(2.51\pm0.48)\times10^{5}$ & 1.22 (304)\\
\hline
\hline
\end{tabular}
\end{center}
\end{table}

\end{document}